\documentclass[twocolumn,superscriptaddress,10pt]{revtex4}

\usepackage{amsmath,amssymb,graphicx}

\begin{document}

\title{Long-lived quantum coherence of two-level spontaneous emission
models within structured environments}

\author{Ping Zhang}

\author{Bo You}
\author{Li-Xiang Cen}
\email{Corresponding author: lixiangcen@scu.edu.cn}
\affiliation{Center of Theoretical Physics, College of Physical Science and
Technology, Sichuan University, Chengdu 610065, China}

\begin{abstract}
We investigate the long-lived quantum coherence of two-level spontaneous
emission models within structured environments. The population of the system
under the asymptotic non-Markovian dynamics is linked to the spectral density
of the reservoir through a general functional relation between them.
We figure out explicitly
the preservation of quantum coherence, via notions
of entanglement and quantum discord, in connection with the spectral parameters
of Ohmic class reservoirs and then show how to achieve them optimally.
We expect these results to contribute towards reservoir
engineering with the aim of enhancing stationary quantum coherence in noisy
environments.
\end{abstract}


\maketitle

Non-Markovian dynamics of open quantum systems attracts intensive attention
recently \cite{breuer2002}. Differing from the conventional Markovian dissipative
process, the memory effects of the non-Markovian environment will lead to
non-exponential decay of the system and even result in dissipationless behavior.
In general, these peculiarities become evident in cases of low-temperature environments
and strong system-reservoir couplings. The corresponding dynamics of the system
will depend dramatically on the spectrum structure of the reservoir.
Besides the fundamental interest to the statistical physics itself,
the study on this subject is stimulated by the progress of quantum
information science \cite{QIT}, since accurate coherence control
of quantum systems under noisy environments requires that the memory
effect of the environment should be taken into account.

Basically, the memory effect of the non-Markovian environment will prolong
quantum coherence, e.g., it may lead to entanglement revival and protect
a composite system against sudden death of entanglement \cite{yu}.
Modifying the property of
the reservoir to reach the non-Markovian regime was shown to be practicable
in some physical systems, including the structured environment of
photonic crystal materials \cite {nature2004,ol2008,PRL2012}
and the optically confined
Bose-Einstein condensate reservoir \cite{bec1}.
For the typical spontaneous emission model, it is known for a
long time that incomplete decay of atomic excitation
can occur in the medium of photonic crystals \cite{photonic1,photonic2}.
The associated phenomenon of entanglement trapping was unveiled
recently \cite{bellomo2008,tong2010,franco2012}. To characterize the
connection between the long-lived quantum coherence and the reservoir spectrum
hence is not only an issue of the non-Markovian dynamics itself, but
also a task for reservoir engineering
to enhance stationary quantum correlations in noisy environments.

Here we focus on the two-level spontaneous emission model and characterize
its longtime behavior in connection with
the reservoir spectra. Interaction of the model
in the rotating-wave approximation is described by
\begin{equation}
H=\omega _0\sigma _{+}\sigma _{-}+\sum_k[\omega _ka_k^{\dagger
}a_k+(g_ka_k\sigma _{+}+g_k^{*}a_k^{\dagger }\sigma _{-})],  \label{hamil1}
\end{equation}
where $\omega_0$ is the fixed frequency of the system, $\sigma _{\pm }$
are shift operators acting on levels $|\pm\rangle $, and $a_k^{\dagger }$ ($a_k$)
are creation (annihilation) operators of bosonic field modes with
frequencies $\omega _k$. In the case that the environment is initially
in the vacuum
state $|0\rangle _E$, evolution of the total system is described by
$|\psi(t)\rangle=c(t)|+\rangle |0\rangle _E +\sum_kc_k(t)|-\rangle |1_k\rangle$,
where
$|1_k\rangle \equiv a_k^\dagger|0\rangle_E$ denotes the single excitation
of the $k$th field mode.
The corresponding Schr\"{o}dinger equation is amenable to an exact resolution
as the spectral function of the reservoir
is given, which indeed has ever been explored intensively, e.g.,
via numerical calculations \cite{numer} and analytical approaches
\cite {photonic1,photonic2,pseudo}. However, since most studies
were focused on the time evolution of the system, explicit revelation
upon the connection between the population and the reservoir spectrum
has only been
obtained for very few cases with particular forms of the spectral density
\cite{photonic2,2012}.

In this work we expose explicitly the link between the long-lived
coherent population of the system and the spectral density of the reservoir
for the spontaneous emission model. By exploiting the
functional relationship between the asymptotic population and
the spectral density, we carry on detailed analyses upon how to engineer
the parameters for Ohmic class spectra to achieve optimally
the stationary coherent population of the asymptotic process.
This enables us to characterize further the trapping phenomenon of quantum
correlations--by notions of entanglement and quantum discord--in
connection with
spectral parameters for a two-qubit system undergoing local dissipative
channels.

Let us start by considering the eigenvalue equation of the Hamiltonian (\ref{hamil1})
in the single-excitation sector. By substituting $|\Phi _{BS}\rangle =b|+,\{0_k\}\rangle
+\sum_kb_k|-,1_k\rangle $ into $H|\Phi _{BS}\rangle =E|\Phi _{BS}
\rangle $, one obtains the secular equation
\begin{equation}
\omega _0-\int_0^\infty \frac{J(\omega )}{\omega -E}d\omega =E,
\label{secular}
\end{equation}
where $J(\omega )=\sum_k|g_k|^2\delta (\omega -\omega _k)$ is the spectral
density of the mode continuum of the reservoir. The corresponding
coefficients of the eigenstate $|\Phi _{BS}\rangle$ are expressed as
$b_k=g_kb/(E-\omega _k)$ and
\begin{equation}
b=\left[ 1+\int_0^\infty \frac{J(\omega )}{(\omega -E)^2}d\omega \right]
^{-1/2}.  \label{bs}
\end{equation}
Since we are considering the coupling between the two-level system and a
continuous spectrum, the solution of Eq. (\ref{secular}) highly depends on
the explicit form of $J(\omega)$. Note that this kind of eigenvalue problem
has ever been investigated in a different physical context \cite{friedrichs,friedrichs1}
and one can even retrospect it to the
work by von Neumann and Wigner \cite{1929}. For the spontaneous emission of
an atom within
photonic band gap mediums, existence of the eigen-solution of the above
equation, so called as atom-photon bound states, was known well and
the associated phenomenon of partial inhibition of radiative decay
of the atomic excited state has been intensively studied in early
works \cite{photonic1,photonic2,numer,pseudo}.

Here, we suppose that Eq. (\ref{secular}) possesses only a single real root,
which is possible if the spectral function fulfills
$J(\omega)>0$ for $\omega>0$ \cite{friedrichs2,miyamoto}. Note that due to
the possible divergence of the integral contained in
Eq. (\ref{secular}), a positive real root $E>0$ must not exist under this
condition. So the real root of Eq. (\ref{secular}), if do exist, should be
unique in view that the left hand of Eq. (\ref{secular}) decreases monotonically
with $E$ in the range $E\in(-\infty,0)$. As a result, for a system
initially in the excited state $|+\rangle$ with the amplitude $c(0)=1$, the residual
population after evolution in the longtime limit is given by
\begin{equation}
P(t)\equiv|c(t)|^2\stackrel{ t\rightarrow \infty }{\longrightarrow }
P_\infty=b^4.  \label{popul}
\end{equation}
To make this result clear, we recall that according to the Schr\"{o}dinger
equation $i\frac {\partial}{\partial t}|\psi (t)\rangle=H|\psi (t)\rangle$,
the time evolution of the
amplitude $c(t)$ satisfies the integro-differential equation \cite{breuer2002}
\begin{equation}
\dot{c}(t)+i\omega _0c(t)+\int_0^tc(\tau )f(t-\tau )d\tau =0,
\label{intdiff}
\end{equation}
with $f(t-\tau )=\int_0^\infty J(\omega )e^{-i\omega (t-\tau )}d\omega $.
In the longtime limit $t\rightarrow \infty$, there will be no ingredient
of the excited state $|+\rangle$ retained
unless Eq. (\ref{secular}) allows the solution of bound states.
Particularly, for the case that Eq. (\ref{secular}) allows only a single
bound state $|\Phi_{BS}\rangle$, one can express the excited state as $|+,\{0_k\}\rangle
=b|\Phi _{BS}\rangle +\bar{b}|\Phi _D(0)\rangle $, where $|\Phi _D(0)\rangle
$ represents the projective state of $|+,\{0_k\}\rangle $ over the
complemental subspace orthogonal to $|\Phi _{BS}\rangle $ and the coefficient
$\bar{ b}=(1-b^2)^{1/2}$. The generated evolution by the
Hamiltonian (\ref{hamil1}), $|\Phi _D(t)\rangle =e^{-iHt}|\Phi _D(0)\rangle $,
will decay entirely to the ground state as
$t\rightarrow \infty $. So the amplitude of the excited state in the longtime
limit is contributed solely by the ingredient of it in $|\Phi_{BS}\rangle $,
which leads promptly to that $|c_\infty |=b^2$. Note that this fact has ever
been displayed in the literature \cite{photonic2} to describe the incomplete
decay of an atom in photonic band gap mediums.

Combination of the above Eqs. (\ref{secular})-(\ref{popul})
suggests a conclusive functional relationship between the long-lived
population and the spectral function: $P_\infty=b^4[\omega_0,J(\omega )]$.
Although each of these expressions, Eqs. (\ref{secular}), (\ref{bs}) and (\ref
{popul}), has ever been obtained previously, say, in literatures \cite
{photonic1,photonic2}, the importance of this functional relation
to explore the long-lived quantum coherence was not revealed so far: it
renders indeed a peculiar perspective to expose the asymptotically
dissipationless
behavior of the system in connection with details of the spectral structure
of the reservoir. This offers straightly the information
towards reservoir engineering, e.g., to the aim of enhancing stationary
quantum coherence in noisy environments.

To proceed, we mention the fact that $P_\infty$ in general is not a
monotonically increasing quantity of the dissipation strength.
This is somewhat counter-intuitive since a strong dissipation strength is
regarded as a necessary condition for the existence of the bound state. To
make it clear, we record $J(\omega)=\eta j(\omega)$ and substitute it
into Eq. (\ref{secular}). One finds that $E$ will descend as
the dissipation strength $\eta$ increases. The variation of $b$ in Eq. (\ref
{bs}), relying on its contained integrand, is determined by the competition
between the numerator and the denominator, both terms increasing with the
strength $\eta$. As a consequence, the monotonicity of $P_\infty$ as a
function of $\eta$ is conditioned to the concrete form of $j(\omega)$.

Let us go ahead by considering the widely used Ohmic class reservoir
with $J(\omega )=\eta \omega _c^{1-s}\omega ^se^{-\omega /\omega _c}$,
where $\omega _c$ is the cutoff frequency and $s$ is a parameter whose
scope, $s<1$, $s=1$, $s>1$, corresponds to sub-Ohmic reservoirs, Ohmic and
super-Ohmic reservoirs, respectively. It turns out that in this case
the solution of Eq. (\ref{secular}) could exist as long as the parameters
satisfy $\eta ^{-1}\leq (\omega _c/\omega _0)\Gamma (s)$,
where $\Gamma (s)$ is the gamma function. The equality contained here is
obtained by substituting directly $E=0$ into the secular equation (\ref{secular}).
This critical condition characterizes actually the occurrence of a quantum
phase transition of the model with or without a bound state
(ground state) \cite{tongprb}.
As the solution of the bound state is unique, the functional relation of
Eqs. (\ref{secular})-(\ref{popul}) yields
\begin{equation}
P_\infty(\eta ,s,\omega _0/\omega _c)=\left[ 1+\int_0^\infty \frac{\eta
x^se^{-x}}{(x-\kappa )^2}dx\right]^{-2},  \label{poh}
\end{equation}
where $\kappa \equiv E/\omega _c$ is determined by
\begin{equation}
\omega _0/\omega _c-\int_0^\infty \frac{\eta x^se^{-x}}{x-\kappa }
dx=\kappa .
\label{trans}
\end{equation}
Numerical calculations to the latter transcendental equation are required to
obtain $\kappa $ for specified parameters $(\eta ,s,\omega _0/\omega _c)$,
with which the exact population $P_\infty$ can be achieved from Eq. (\ref{poh}).

The dependence of $P_\infty$ on spectral parameters of the reservoir
determined by the implicit function of Eq. (\ref{poh}) is quite sophisticated.
We present below a detailed
analysis on how to achieve $P_\infty$ optimally with respect to different
zones of the spectral parameters $(\eta ,s,\omega _0/\omega _c)$. In the
case of low $\omega_c/\omega_0$, a high value of $\eta\Gamma(s)$ indicates
that the scope of the parameters $(\eta,s)$ is relatively narrow to achieve
a nonvanishing $P_\infty$. Note that there is a physical constraint of the
coupling strength in order to validate the rotating-wave approximation for
the model Hamiltonian (\ref{hamil1}) (reasonably $\eta \lesssim 0.1$ owing
to $|g_k|\ll \omega_0$). The dependence of $P_\infty$ on $\eta$ is shown in
Fig. 1(a) with $\omega_c/\omega_0=0.3$. Our calculation displays that high
Ohmicity about $s\gtrsim 5.25$ is required, which may challenge the
technology of the reservoir engineering. For the situation with high cutoff
frequencies, the dependence of the population $P_\infty$ on the parameter $s$
is shown in Fig. 1(b). Note that in the limit of $\omega _0/\omega
_c\rightarrow 0$, the transcendental equation defines an implicit function
$\kappa =\kappa (\eta ,s)$, hence $P_\infty=P_\infty(\eta ,s)$ according to
Eq. (\ref{poh}). For $\eta=0.08$, the maximum of the asymptotic population
$P_\infty\simeq0.9$ is achieved at $s\simeq2.34$.

\begin{figure}[h]
\begin{center}
\includegraphics[width=8cm,trim=0 60 0 60]{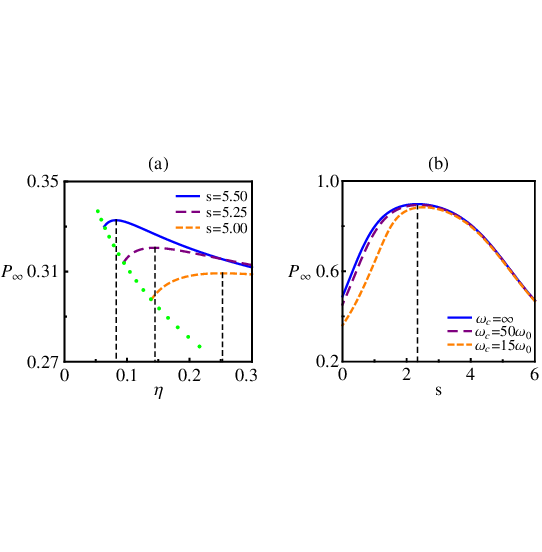}
\end{center}
\caption{Long-lived population $P_\infty=b^4$ in relation to the parameters
of Ohmic class spectra. (a) $P_\infty$ as a function of $\eta$ with $
\omega_c=0.3\omega_0$, $s=5$, $5.25$, and $5.5$, respectively. The
green-dotted line figures out the boundary determined by the critical
condition $\eta\Gamma (s)=\omega_0/\omega_c$. For $s=5.5$, the maximal $
P_\infty=0.33$ is achieved at $\eta=0.08$. (b) $P_\infty$ as a function of $s
$ with $\eta=0.08$, $\omega_c/\omega_0=15$, $50$ and $\infty$. In the limit $
\omega_c/\omega_0\rightarrow \infty$, the maximal $P_\infty=0.90$ is
achieved at $s=2.34$.}
\end{figure}

As an application of the above results, we investigate the long-lived
behavior of quantum correlations of an initially entangled two-qubit state
$\rho_{AB}$, in which the subsystem $A$ is subject to a local channel of
spontaneous emission. Referring to different notions, quantum correlation
of the system is often depicted either by entanglement or quantum discord.
The amount of entanglement of the two-qubit state $\rho_{AB}$ can be
expressed explicitly via the measure of concurrence \cite{wootters}:
$C_{AB}=\max\{0,\lambda_1^{1/2}
-\sum_{i=2}^4\lambda_i^{1/2}\}$, where $\lambda_i$ are eigenvalues of the
matrix $\rho_{AB}\sigma_y\otimes \sigma _y\rho_{AB}^*\sigma_y\otimes\sigma_y$
in descending order. The discord of $\rho_{AB}$, according to the definition of
\cite{discord1,discord2}, is given by $Q_{AB}=S(\rho _A)-S(\rho _{AB})+
\min_{\{A_k\}}\sum_kp_kS(\rho_B^k|\{A_k\})$, where $S(\rho)\equiv-\mathrm{{tr}
(\rho \log_2\rho)}$ is the von Neumann entropy. To describe the evolution of
$\rho_{AB}$, we note that the damping channel of the qubit $A$ undergoing
the spontaneous emission can be depicted by the Kraus representation
$\rho_A(t)=\sum_{i=1}^2\Gamma _i(t)\rho_A(0)\Gamma_i^{\dagger}(t)$, in which
\begin{equation}
\Gamma _1(t)=\left[
\begin{array}{ll}
0 & 0 \\
\bar{c}(t) & 0
\end{array}
\right] ,~~\Gamma _2(t)=\left[
\begin{array}{ll}
c(t) & 0 \\
0 & 1
\end{array}
\right]  \label{kraus}
\end{equation}
with $\bar{c}(t)\equiv[1-|c(t)|^2]^{1/2}$. Therefore one has
\begin{equation}
\rho _{AB}(t)=\sum_{i=1}^2\Gamma _i(t)\otimes I_B\rho _{AB}(0)\Gamma
_i^{\dagger }(t)\otimes I_B.  \label{2qubit}
\end{equation}

We characterize below quantum correlations for the case of pure
input states: $|\psi_{AB}(0)\rangle =\alpha|+-\rangle +\beta|-+\rangle $
with $|\alpha|^2+|\beta|^2=1$.
The concurrence of the corresponding $\rho_{AB}(t)$
is obtained readily as $C_{AB}(t)=2|\alpha \beta c(t)|$. Also the discord can
be worked out via the method in \cite{cen}, expressed analytically as:
$Q_{AB}(t)=h(\lambda )+h(\lambda _A)-h(\lambda _{AB})$,
where $h(x)=-x\log _2x-(1-x)\log _2(1-x)$ is the binary entropy function and
the parameters
$\lambda _A=|\alpha c(t)|^2$, $\lambda _{AB}=|\alpha \bar{c}(t)|^2$,
and $\lambda =\frac 12\{1+[1-4|\alpha\beta\bar{c}(t)|^2]^{1/2}\}$.
In the limit $t\rightarrow \infty$, a steady population
$|c_\infty|^2=b^4[J(\omega)]$ is yielded, hence
there are $Q_{AB}^\infty=Q_{AB}^\infty[J(\omega)]$
and $C_{AB}^\infty=C_{AB}^\infty[J(\omega)]$.
We depict in Fig. 2 the two quantities varying with the parameter $s$ of
the Ohmic class spectra with $\eta=0.08$ and $\omega_c\gg\omega_0$.
Since both the two quantities of the output state are monotonic
functions of the amplitude $|c_\infty|$, the maximal values of
$C_{AB}^\infty$ and $Q_{AB}^\infty$ are obtained at the same point
with $s\simeq 2.34$.

\begin{figure}[h]
\begin{center}
\includegraphics[width=8cm,trim=0 60 0 60]{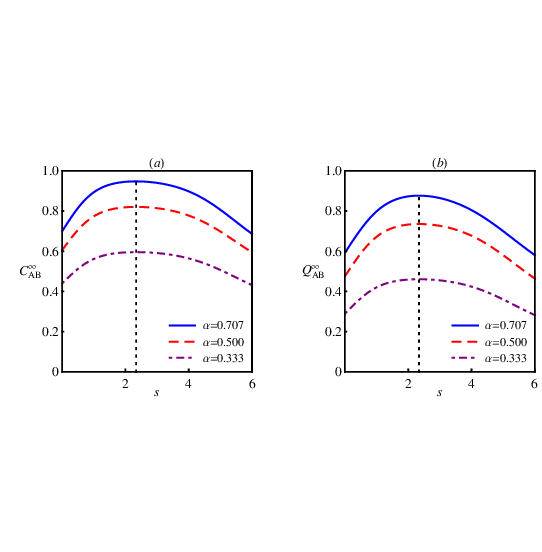}
\end{center}
\caption{Long-lived entanglement and quantum discord as a function of $s$
for the Ohmic class reservoirs with $\eta=0.08$ and $\omega_c\gg \omega_0$.
The two-qubit system is initially in a pure state
$|\psi_{AB}(0)\rangle =\alpha|+-\rangle +\beta|-+\rangle $. The maximal values
$C_{AB}^\infty\simeq0.95$ and $Q_{AB}^\infty\simeq0.88$ are obtained at $s\simeq2.34$
for the maximally entangled input state.}
\end{figure}

To summarize, we have studied the asymptotic behavior of two-level
spontaneous emission models within structured environments.
As the occurrence of long-lived quantum coherence is clearly
a consequence of the memory effect of the non-Markovian dynamics,
our calculations reveal that it can occur for a wide range of
Ohmic class reservoirs. We expect that our derived results,
figuring out explicitly the connection between the system behavior and the
reservoir spectra, could contribute useful information towards reservoir
engineering to enhance stationary quantum correlations under noisy
environments. Finally, we mention that we have assumed a model with
a bosonic reservoir in the vacuum state initially. Further studies
to explore the influence of the existence of the bound state upon the
coherence of the system dynamics for a realistic environment with nonzero
temperature should be a subject of future researches.

We acknowledge Paolo Zanardi and Jiushu Shao for helpful discussions. This
work was supported by the Natural Science Foundation of China.

\end{document}